\documentclass[3p,times,twocolumn]{elsarticle}

\usepackage{ecrc}

\volume{00}

\firstpage{1}

\journalname{Nuclear and Particle Physics Proceedings}

\runauth{}

\jid{nppp}

\jnltitlelogo{Nuclear and Particle Physics Proceedings}

\usepackage{amssymb}

\usepackage[figuresright]{rotating}

\begin{document}

\begin{frontmatter}



\dochead{}

\title{Productions of $\eta$, $\rho^0$ and $\phi$ at large transverse momentum in Heavy ion Collisions}

\author[label1,label2]{Wei Dai}
\author[label2]{Ben-Wei Zhang}

\address[label1]{Physics Department, Tsinghua University, Beijing, China}
\address[label2]{Key Laboratory of Quark \& Lepton Physics (MOE) and Institute of Particle Physics,
 Central China Normal University, Wuhan 430079, China}

\begin{abstract}

The suppression of the productions of the $\eta$ meson in relativistic heavy-ion collisions and its ratio of $\eta/\pi^0$ are computed theoretically in the framework of the perturbative QCD(pQCD) to confront the experimental data which matches well. We explore how the hadron production ratios as $\eta/\pi^0$ would further disclose the informations of the production suppressions due to the energy loss of the energetic jet that propagating though the QGP medium. Also, we present our further studies on vector mesons such as $\rho^0$ and $\phi$ within the same framework. The theoretical predictions based on pQCD are thus firstly given which give a decent description on the experimental measurements. It paved the way to the uniformly understanding of the strong suppression of single hadron productions at large transverse momentum which is a convincing evidence of the jet quenching effect.
\end{abstract}


\end{frontmatter}


\section{Introduction}
\label{Introduction}

The so called quark-gluon plasma(QGP) is produced at the earlier stage after the hard scattering in the heavy ion collisions experiments at  various ultra relativistic energies.  The properties of the de-confined state of quarks and gluons thus can be reflected by the modifications of the final state measurements relative to the ones in $p+p$ collision. One of the important properties is the jet quenching effect~\cite{Wang:1991xy} which indicate that the energetic partons produced in the hard scattering will lose their energies when propagating in the hot and dense medium. Thanks to the developments in both experiments and theories, besides the hadron production suppression, there are more hard probes to help constraint the jet quenching models from RHIC to LHC energies such as leading hadron yield at large $p_T$ and the related correlations,   and full jet observables~\cite{Gyulassy:2003mc,Vitev:2008rz,Vitev:2008bx,Dai:2012am,Dai:2013xca,Dai:2015dxa}. Even though the hadron suppression is the earliest evidence and well measured observable, it is still playing an important role when understanding the jet quenching phenomena, and a lot of challenging questions are still waiting for us to explore.  Firstly, we need to know whether the jet quenching description can explain most of the identified hadrons production suppressions simultaneously.  Unlike $\pi$ meson, the other identified hadrons such as $\eta$, $\phi$, $\rho$, the yields are not rich compared to $\pi$ meson, and the constrainst of their fragmentation functions are rather loose or even not available due to the lack of the experimental data. Secondly, the interplay between theory and experiment of the hadron production ratio such as $\eta/\pi^0$ also help understanding how the flavor dependent parton energy loss alter the flavor compositions of fast partons~\cite{Liu:2006sf, Chen:2008vha}.


In the talk, we firstly explore the $\eta$ meson production in $A+A$ collisions, also its suppression pattern and production ratio to $\pi^0$, since $\eta$ is the second important source of decay electrons and photons just after $\pi^0$. We calculate the $\eta$ productions and its suppression in heavy ion collisions at the RHIC and LHC using the higher twist approach which considering multiple scattering of the fast parton traversing through the dense QCD matter. The features of $\eta/\pi^{0}$ ratios in both $\rm p+p$ and $\rm A+A$ collisions are also explored. In the same framework, a newly developed initial parameterizations of $\rho^{0}$ and $\phi$ meson fragmentation functions at a starting energy scale of $\rm Q_{0}^2=1.5(GeV)^2$~\cite{Saveetha:2013jda,Indumathi:2011vn} are also employed to give a first prediction of $\rho^{0}$ and $\phi$ meson productions in $\rm p+p$ collisions when having them evolved through DGLAP evolution equations at NLO~\cite{Hirai:2011si}.  Their nuclear modification factors are also predicted.


\section{Formalism in $p+p$ Collisions}
\label{pp}
The formula of identified hadron $p_{T}$ yield in $p+p$ collisions could be written as:
\begin{eqnarray}
\frac{1}{p_{T}}\frac{d\sigma_{\pi^{0}, \eta}}{dp_{T}}=\int f_{q}(\frac{p_{T}}{z_{h}})\cdot D_{q\to \eta, \pi^{0}}(z_{h}, p_{T})\frac{dz_{h}}{z_{h}^2} \nonumber  \\
+ \int f_{g}(\frac{p_{T}}{z_{h}})\cdot D_{g\to \eta, \pi^{0}}(z_{h}, p_{T})\frac{dz_{h}}{z_{h}^2}  \,\,\, .
\label{eq:ptspec}
\end{eqnarray}
The above equation shows that the hadron yield in $\rm p+p$ will be determined by two factors: the initial (parton-)jet spectrum $f_{q,g}(p_T)$ and the parton fragmentation functions $D_{q,g\to \eta,\pi^{0}}(z_{h}, p_{T})$.  The non-perturbative input of $\eta$ FFs provided by AESSS~\cite{Aidala:2010bn} help us to calculate the single-inclusive $\eta$ meson production as a function of final state $p_T$ in hadron-hadron collisions at NLO. We compare our calculation with PHENIX data~\cite{Adare:2010cy}, and it is observed that the computed inclusive $\eta$ spectrum in $p+p$ collisions with the scale $\mu=1.0 p_T$ agrees well with the PHENIX data.

\section{Suppression Pattern in $A+A$ Collisions}
\label{AA}
\begin{figure}[!t]
\begin{center}
\hspace*{-0.1in}
\includegraphics[width=1.5in,height=1.4in,angle=0]{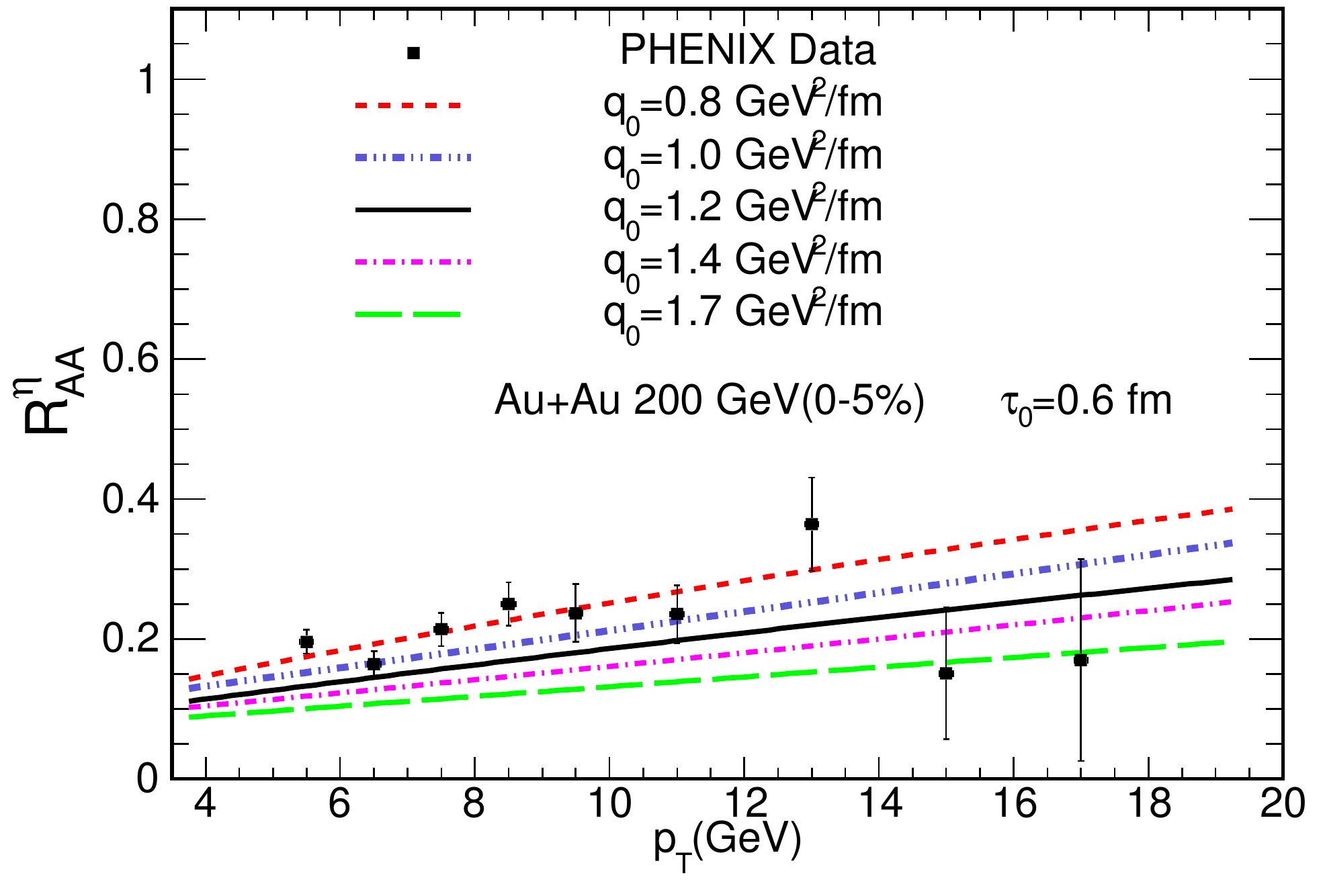}
\includegraphics[width=1.5in,height=1.4in,angle=0]{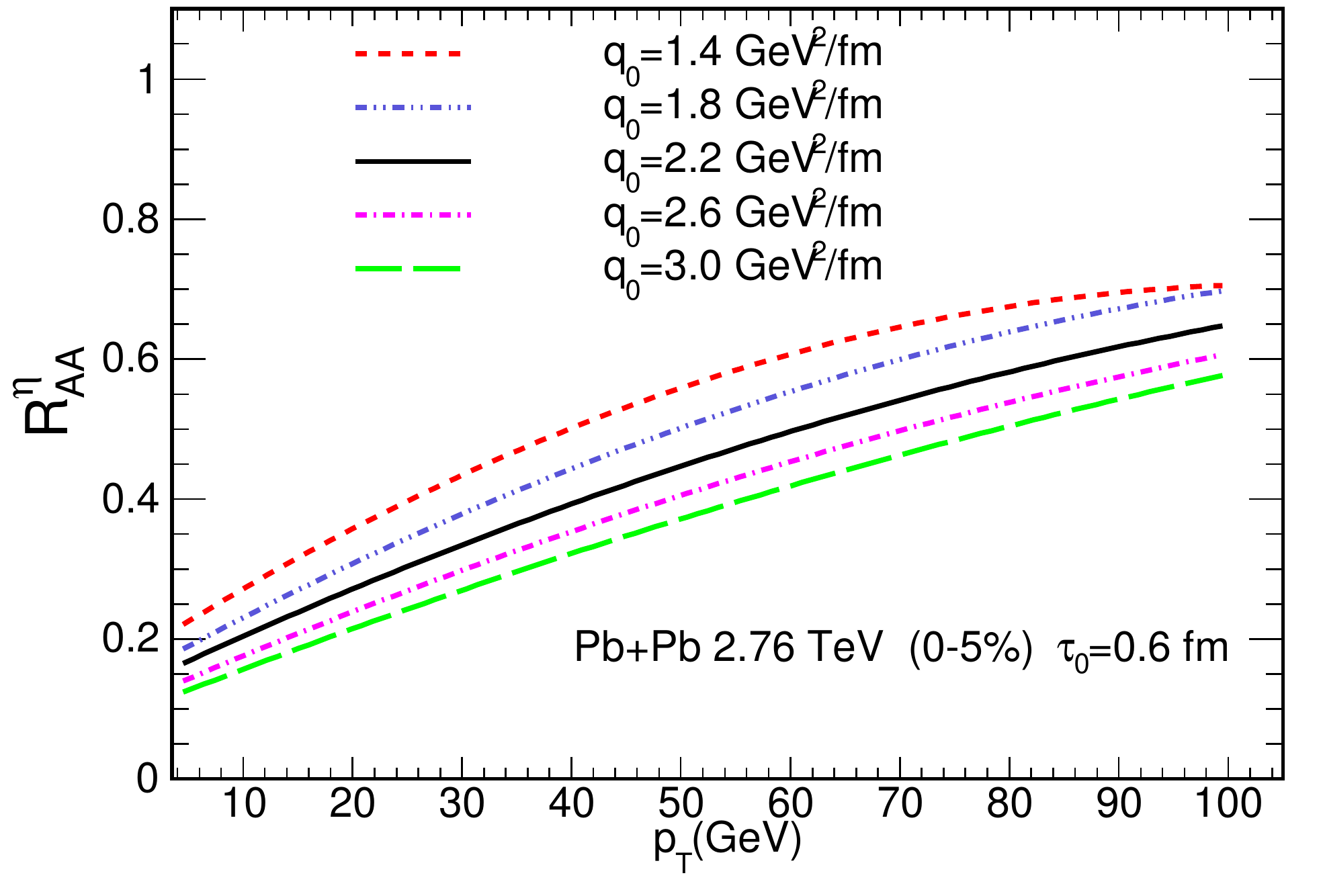}
\hspace*{-0.1in}
\caption{Left panel: comparison between the PHENIX Data of $\eta$ nuclear modification factor in $\rm{Au+Au}$ collisions at
$200$~GeV and numerical simulations at NLO; Right panel: predictions of $\eta$ medium modification factor in $\rm{Pb+Pb}$ collisions at
$2.76$~TeV with different $\hat{q}_0$ parameters.}
\label{fig:illustetaraarhic}
\end{center}
\end{figure}

The resulting parton energy loss due to multiple scattering leads to effectively modified parton fragmentation functions in medium (mFF) which is calculated with generalized QCD factorization of twist-4 processes~\cite{Gyulassy:2003mc}. Incorporating with next- to-leading order (NLO) pQCD improved parton model, the phenomenological study on $\pi^0$ suppression in nucleus-nucleus collisions~\cite{Chen:2010te,Chen:2011vt} gives a fairly good description of $\pi^0$ yields in Au + Au collisions at the RHIC and in Pb + Pb reactions at the LHC. The same model to investigate $p_T$ spectrum of $\eta$ production at NLO in HIC, with the same jet transport parameters $ \hat{q} $ extracted in $\pi^0$ production in HIC~\cite{Dai:2015dxa}. At various values of $\hat q_{0} \tau_{0}=0.48-1.02$ GeV$^2$,we calculate the $Au+Au$ productions within medium modified $\eta$ fragmentation functions in the $5\%$ most central $Au+Au$ collisions at RHIC energy $\sqrt{s}=200$~GeV, and compared our results with  the PHENIX experimental data on $R_{AA}$ of the $\eta$ spectra. The theoretical calculation results explained the data well in large $p_T$ region in the left panel of Fig.~\ref{fig:illustetaraarhic}. The well agreement with the PHENIX data~\cite{Adler:2006hu} at mid-rapidity in the range $p_T = 2-20$~GeV shows that, even $\eta$ meson is 4 times heavier than $\pi^0$, a similar flat production suppression has been observed at RHIC in this $p_{T}$ range independent of their mass. The prediction of $R_{AA}$ of the $\eta$ spectra in the central $Pb+Pb$ collisions at the LHC energy $\sqrt{s}$~=~$2.76$~TeV is given in the right panel of Fig.~\ref{fig:illustetaraarhic}. The $\hat q_{0}\tau_{0}$ are chosen from $0.84-1.8$ GeV$^{2}$.

\hspace{0.7in}
\hspace*{-0.5in}

\section{$\eta/\pi^0$ in $A+A$ Collisions}
\label{ratio}
We also explore the features of $\eta/\pi^{0}$ ratios in both $\rm p+p$ and $\rm A+A$ collisions. We plot the $p_T$ dependence of the $\eta/\pi^0$ ratios in $Au+Au$ at $200$~GeV as in the left panel of Fig.~\ref{fig:illustetapirhic}, and a good agreement between the model calculations with PHENIX data can be seen.  We also predict the $p_T$ dependence of the $\eta/\pi^0$ ratios in $Pb+Pb$ at $2.76$~TeV in the right panel of Fig.~\ref{fig:illustetapirhic}.  Similar trend could be seen at the RHIC and LHC that with the increasing of $p_{T}$, the $\eta/\pi^0$ ratio in $A+A$ collisions comes closer to the $p+p$ curve, and at very larger $p_T$, two curves coincide with each other.

\hspace{0.7in}
\begin{figure}[!t]
\begin{center}
\hspace*{-0.1in}
\includegraphics[width=1.5in,height=1.4in,angle=0]{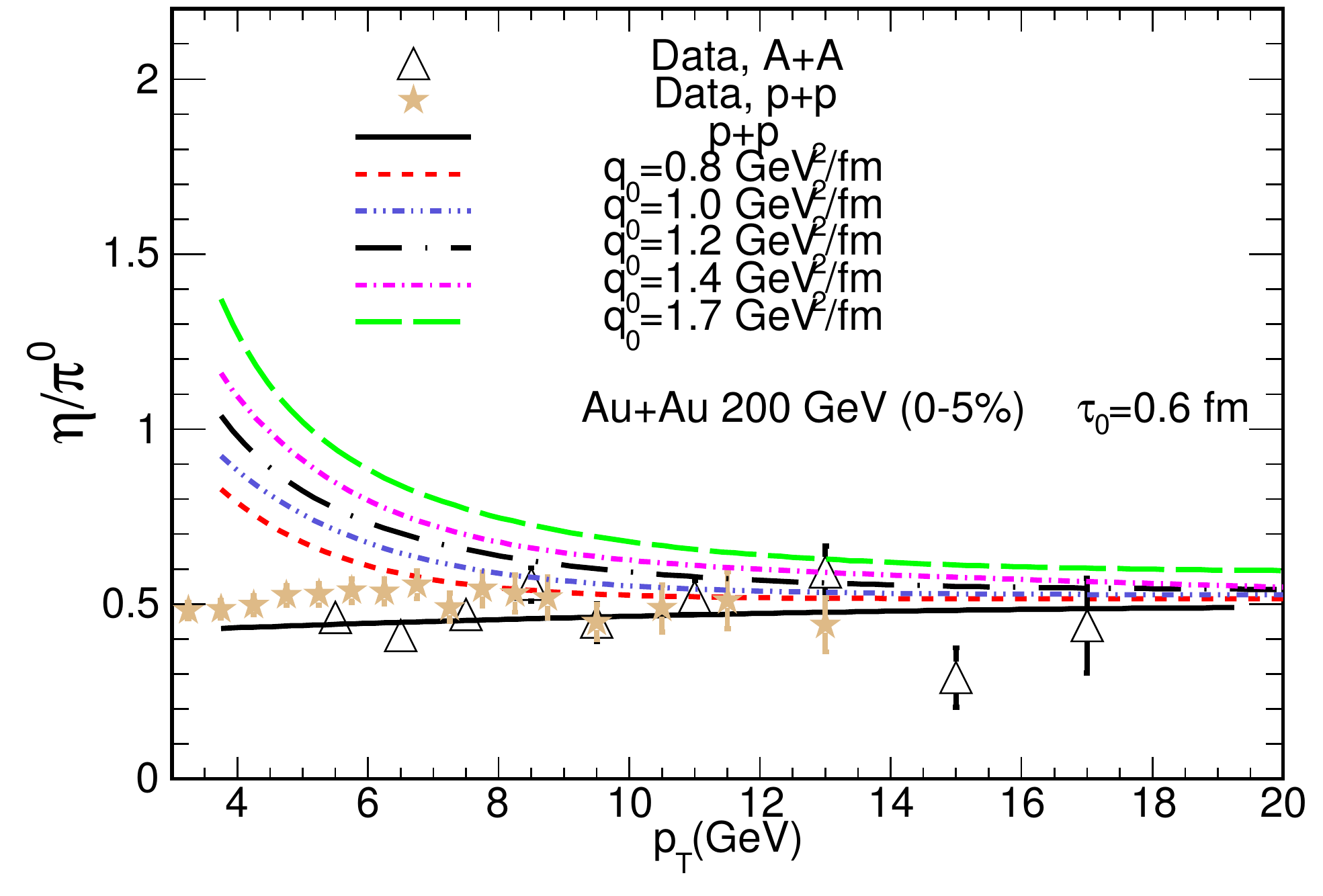}
\includegraphics[width=1.5in,height=1.4in,angle=0]{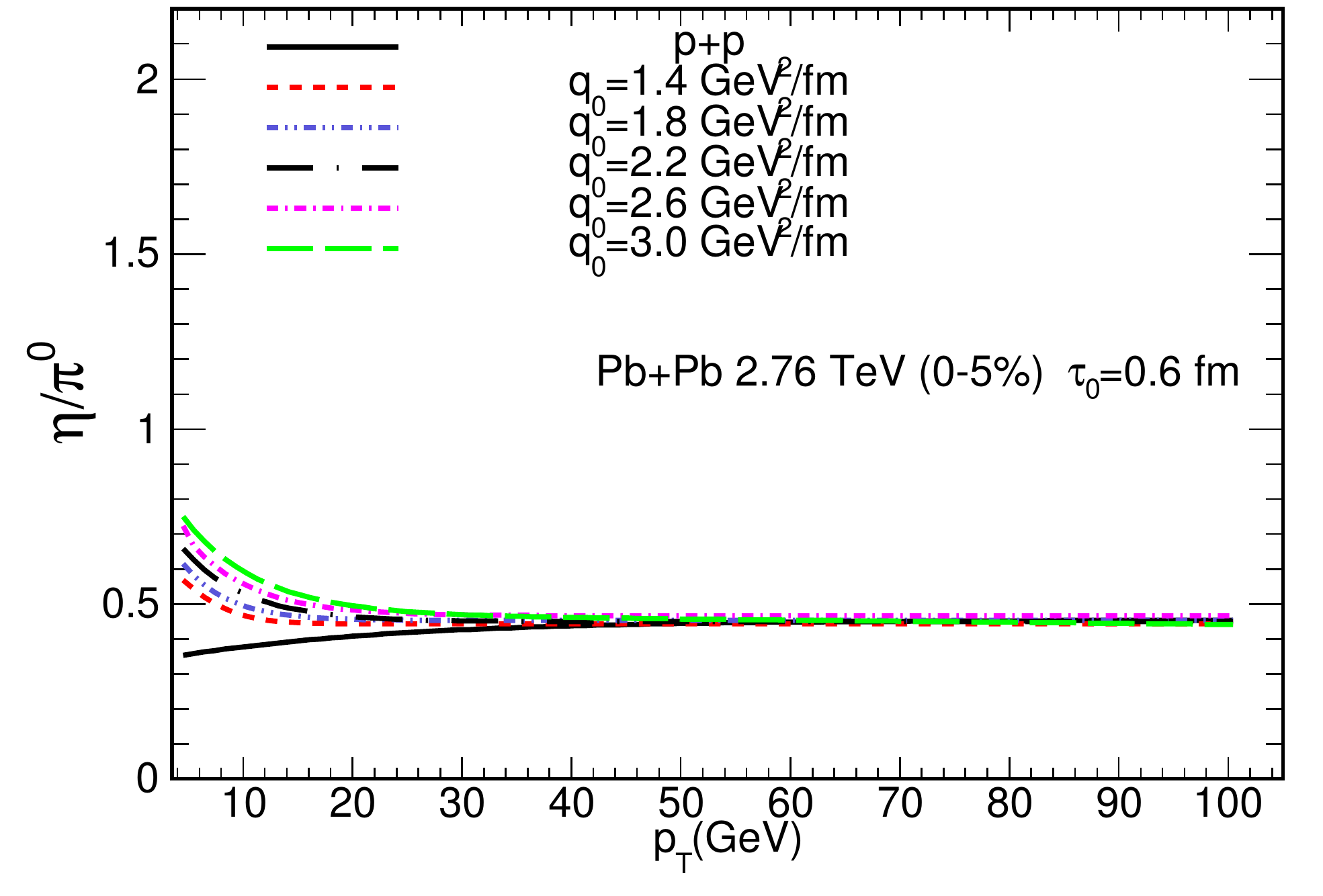}
\hspace*{-0.1in}
\caption{Left panel: comparison between the PHENIX data of $\eta/\pi^0$ ratio in$\rm{p+p}$ collisions and $\rm{Au+Au}$ collisions at
$200$~GeV and the numerical simulations at NLO. Right panel: the NLO pQCD theoretical prediction of $\eta$ /$\pi^0$ ratio in $\rm{Pb+Pb}$ collisions at
$2.76$~TeV.
}
\label{fig:illustetapirhic}
\end{center}
\end{figure}
\hspace*{-1.5in}

The flavor dependent parton energy loss in the QGP may alter the flavor compositions of fast partons, in QCD medium gluon jet suffers larger energy loss than quark jet. Because quark FFs and gluon FF into $\eta$ and $\pi$ have quite different features, in principle, a change of flavor compositions of parton jets may affect the ratio of $\eta/\pi^0$~\cite{Wang:1998bha}.  The coincidence of the $A+A$ and the $p+p$  $\eta/\pi^{0}$ curves in a wide $p_T$ region can not be explained in one simple story that parton jets loss their energies first in the QCD medium and then fragment into hadrons in the vacuum~\cite{Adler:2006hu}.

At very high $p_T$, quark FF is much larger than gluon FF for both $\eta$ and $\pi^{0}$ and shows weak $p_{T}$ and $z_{h}$ dependence in the typical $z_h$ region $0.4-0.7$~\cite{Dai:2015dxa}. The yields of both $\pi^0$ and $\eta$ should also predominantly come from quarks shown in right top panel of Fig.~\ref{fig:illustetapirhic}, and the ratio $\eta/\pi^0$ will also be determined only by quark FFs in vacuum with $z_h$ shift due to energy loss effect. It explain why the ratios of $\eta/\pi^0$ in both $\rm A+A$ and $\rm p+p$ should overlap with the one in $e^+e^-$ scattering, and reach a universal value $\sim 0.5$.

The calculation reflected by the left bottom panel of Fig.~\ref{fig:illustetapirhic}  implies that whereas $\pi^0$ and $\eta$ yields in $\rm p+p$ come from quark or gluon hadronization, when calculating the ratio $\eta/\pi^0$, because of the relative identical fractional contributions of gluon and quark to $\pi^0$ and $\eta$, we can consider the contributions of only quarks (or gluons). The flavor compositions or mixture of quarks and gluons in $\rm p+p$ have nearly negligible effect on $\eta/\pi^0$. Also, it can minimized the effect of the flavor dependent energy loss in hot QCD medium, which underlies the overlapping of $\eta/\pi^{0}$ in $\rm A+A$ and the one in $\rm p+p$ in a wide region of $p_T$. 

Therefore, there are small modifications to $\eta/\pi^0$ due to jet quenching effect when it comes lower $p_T$ region. When plotting the gluon and quark contribution fractions to $\eta$ and $\pi^{0}$ yields in $Au+Au$ collisions at $200$~GeV in the right panel of Fig.~\ref{fig:illustetapirhic}.  It disclosed that the flavor dependent jet energy loss can results in the suppression of gluon contribution fraction in heavy-ion collisions as compared to $\rm p+p$. Previously when we naively expect that because of gluon may give larger $\eta/\pi^0$ ratio than quark does, the larger suppression of gluons in the QCD medium will reduce $\eta/\pi^0$. However, the fact is contrary,  the right panel of Fig.~\ref{fig:illustetapirhic} indicate that instead the jet quenching effect will enhance the ratio $\eta/\pi^0$ in high-energy nucleus-nucleus collisions. The right top panel of Fig.~\ref{fig:illustetapirhic} gives us the reason, the suppression of gluon in QCD medium imposes a larger reduction of the yield of $\pi^{0}$ than that of $\eta$ which gives rise to a slightly enhanced $\eta/\pi^0$ ratio in $\rm A+A$. The  bottom panel here demonstrate a larger  $\eta/\pi^0$ ratio with only gluons is included in the theoretical simulations. 

Three factors of the identified hadron yield in heavy-ion collisions has to be taken into account when understanding any hadron production ratios such as $\eta/\pi^0$: the initial hard jet spectrum, the energy loss mechanism, and parton fragmentation functions to the hadron in vacuum. The simple estimation with only flavor dependent energy loss picture and fragmenting parton components in vacuum may sometimes lead to a wrong conclusion.

\hspace{0.7in}
\begin{figure}[!t]
\begin{center}
\hspace*{-0.1in}
\includegraphics[width=1.5in,height=1.7in,angle=0]{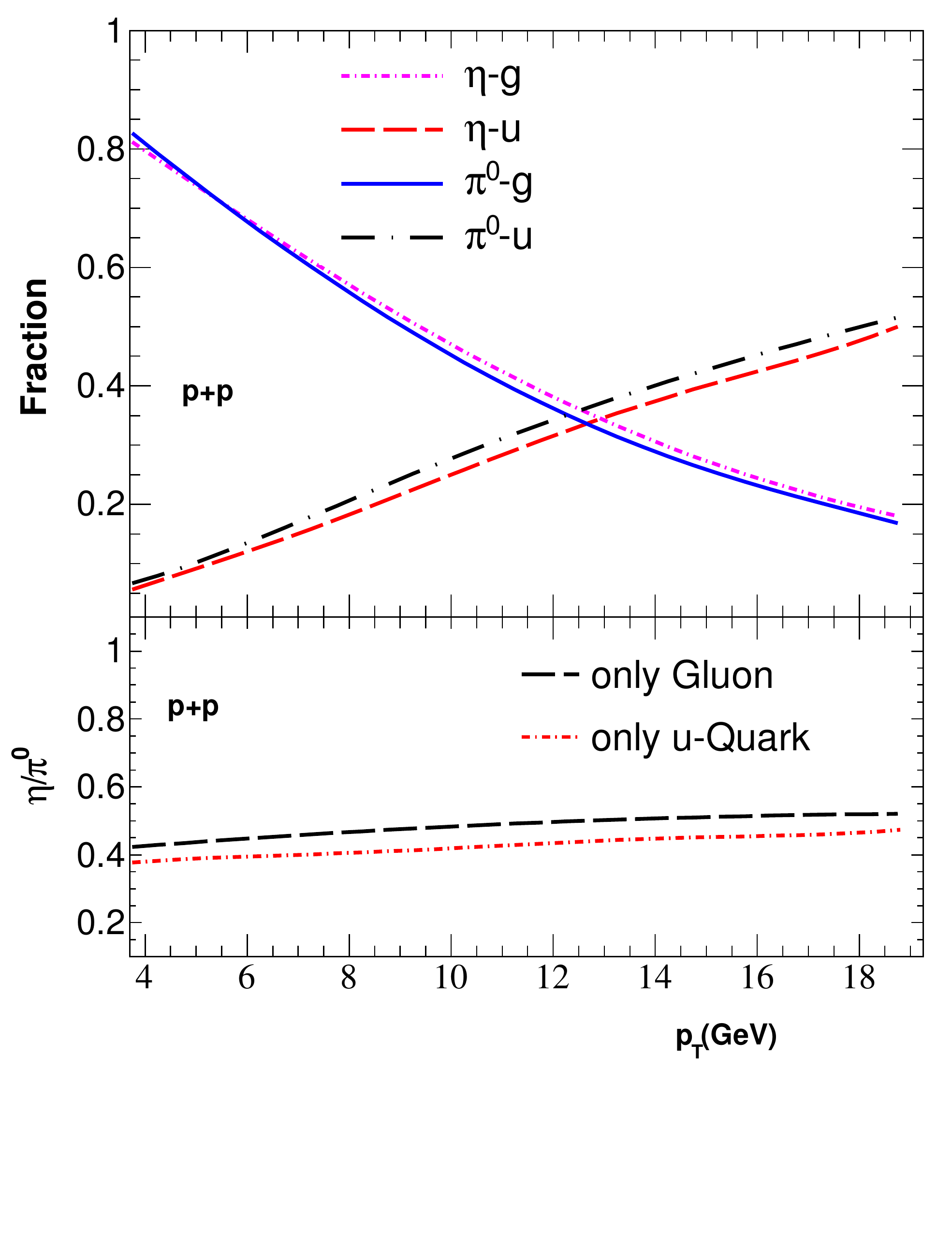}
\includegraphics[width=1.5in,height=1.7in,angle=0]{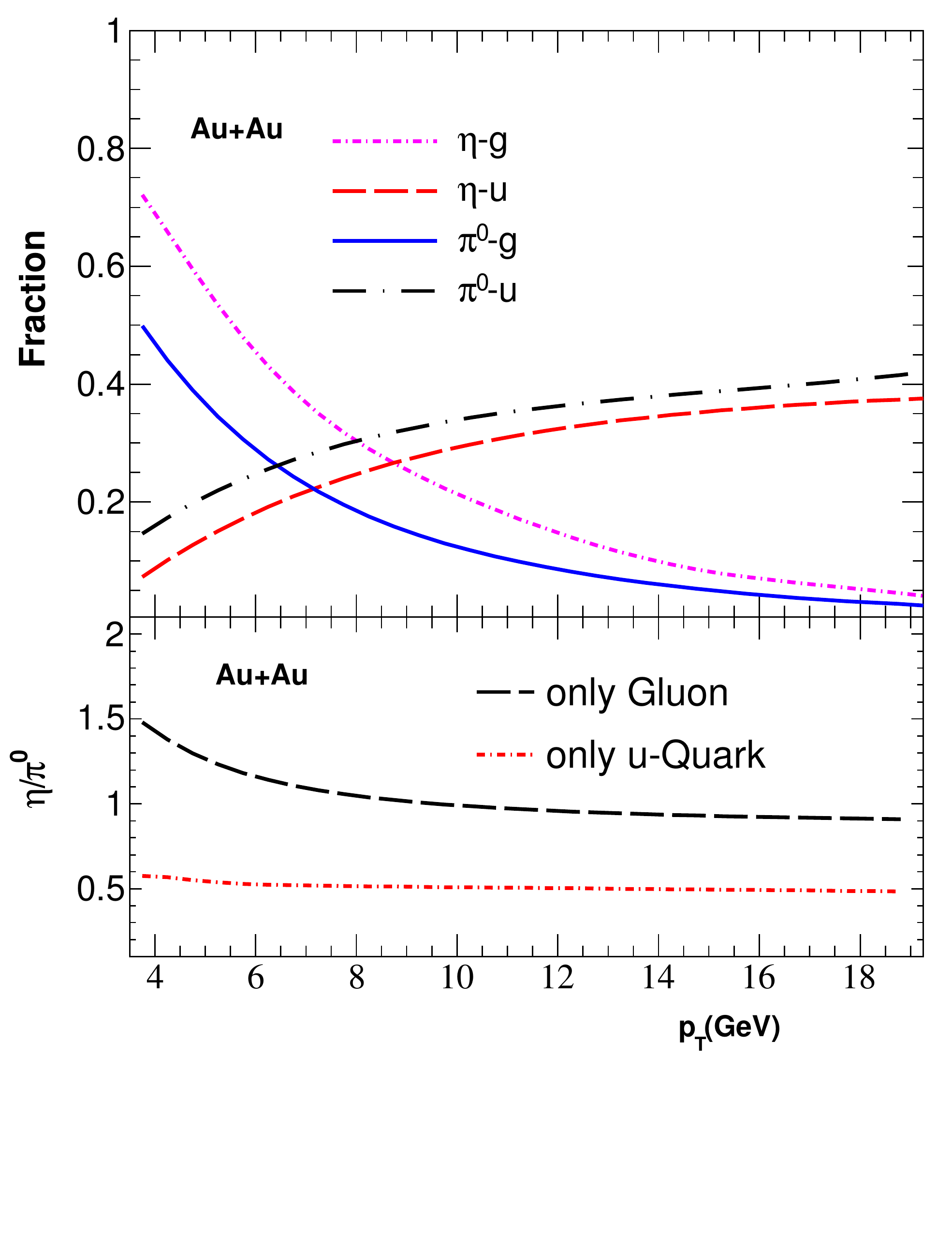}
\hspace*{-0.1in}
\caption{Left top panel: quark and gluon contribution fractions to total $\pi^{0}$ (or $\eta$ meson) yields at NLO in $\rm p+p$ collisions at $200$~GeV.
Left bottom panel: the ratio $\eta/\pi^{0}$ at NLO when only gluon or up quark contribution is considered in $\rm p+p$ collisions at $200$~GeV. Right top panel: quark and gluon contribution fractions to total $\pi^{0}$ (or $\eta$ meson ) yields  at NLO in $\rm Au+Au$ collisions at $200$~GeV;
Right bottom panel: the ratio $\eta/\pi^{0}$ at NLO when only gluon or up quark contribution is considered in $\rm Au+Au$ collisions at $200$~GeV
}
\label{fig:illustetapirhic}
\end{center}
\end{figure}
\hspace*{-1.5in}

\section{$\rho^0$ and $\phi$ Meson Productions in $A+A$ Collisions}
\label{meson}
\begin{figure}[!t]
\begin{center}
\hspace*{-0.1in}
\includegraphics[width=2.4in,height=1.9in,angle=0]{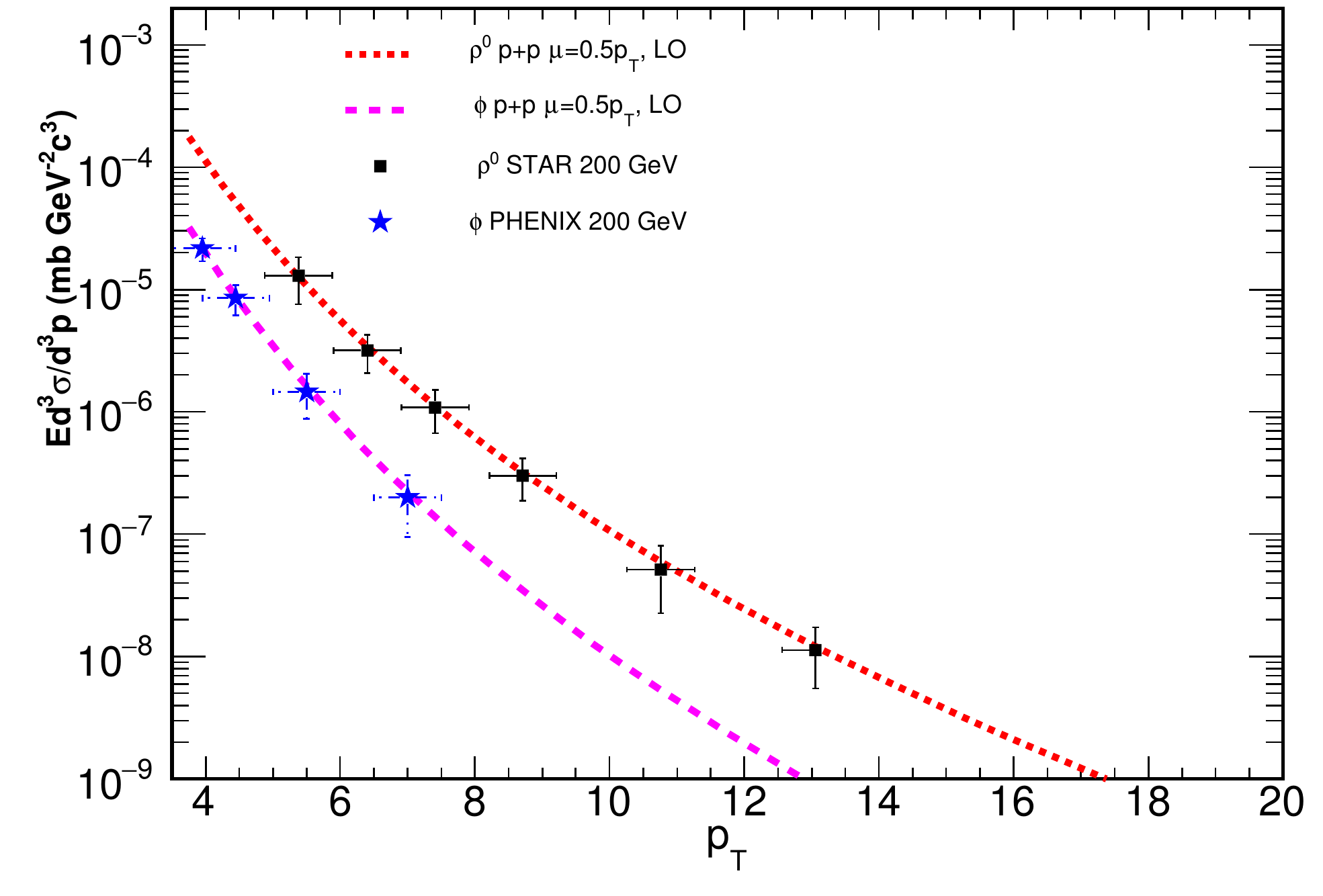}
\hspace*{-0.1in}
\vspace*{-0.2in}
\caption{ Numerical calculation of the $\phi$ and $\rho^{0}$ production in $\rm p+p$ collisions at RHIC $200$~GeV comparing with STAR~\cite{Agakishiev:2011dc} and PHENIX~\cite{Adare:2010pt} data
respectively.}
\label{fig:illustrhopp}
\end{center}
\end{figure}

Now we discuss the numerical predictions of $\rho^0$ and $\phi$ meson productions in the same theoretical framework. Since the lack of experimental data, the newly developed initial parameterizations of $\rho^{0}$ and $\phi$ meson fragmentation functions at a starting energy scale of $\rm Q_{0}^2=1.5(GeV)^2$~\cite{Saveetha:2013jda,Indumathi:2011vn} is only fitted with the LO calculation. Have them evolved through DGLAP evolution equations at LO~\cite{Hirai:2011si}, the $\rho^{0}$ and $\phi$ meson productions in $\rm p+p$ collisions are firstly obtained using pQCD improved parton model,  we find that they describe the experimental data well as shown in Fig.~\ref{fig:illustrhopp}. 
\hspace{0.7in}
\hspace*{-0.5in}

\begin{figure}[!t]
\begin{center}
\hspace*{-0.1in}
\vspace*{-0.1in}
\includegraphics[width=1.5in,height=1.4in,angle=0]{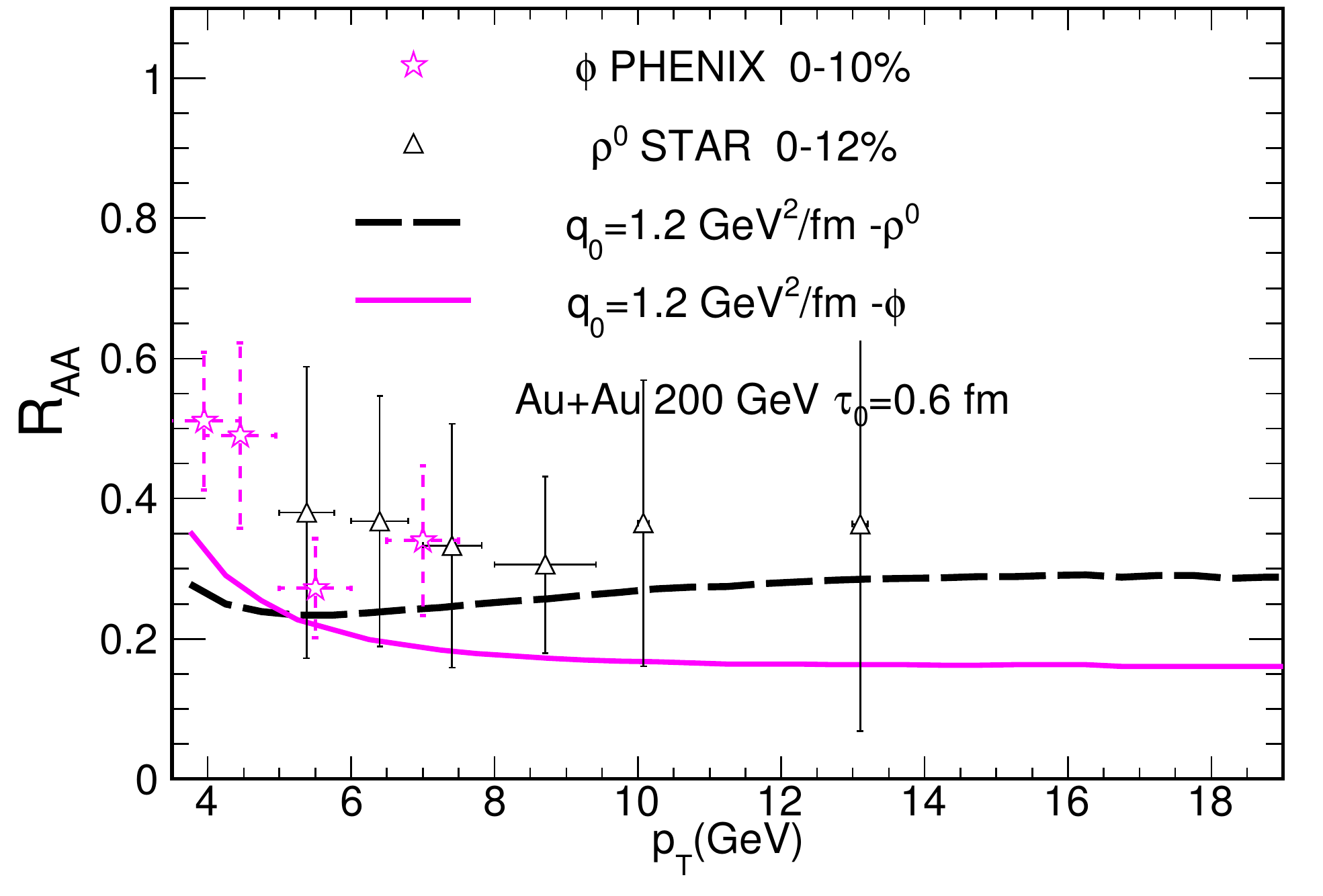}
\includegraphics[width=1.5in,height=1.4in,angle=0]{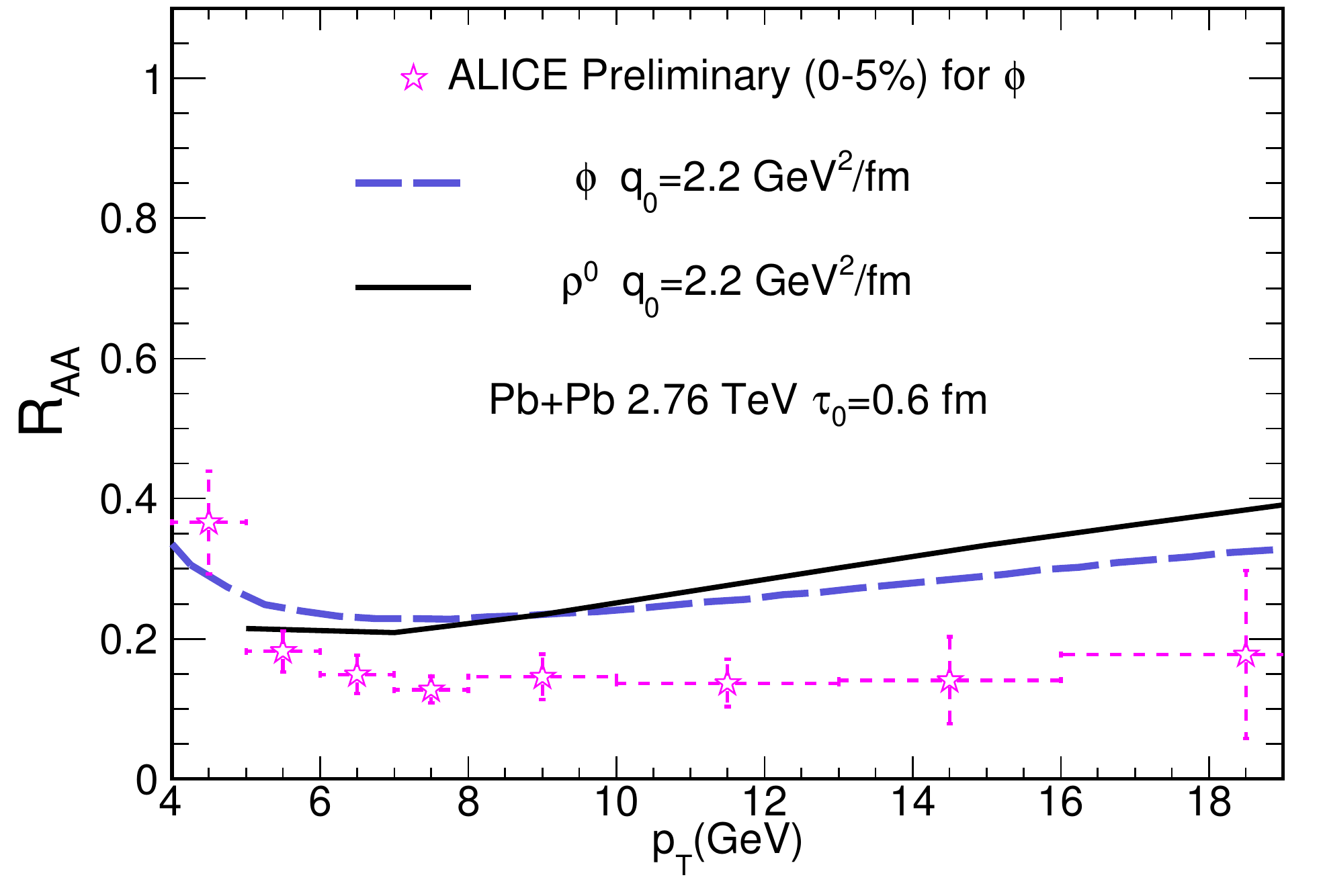}
\hspace*{-0.1in}
\vspace*{0.0in}
\caption{Left panel: numerical calculation of the $\phi$ and $\rho^{0}$ production in $0-10\%$ $\rm Au+Au$ collisions at RHIC $200$~GeV comparing with STAR~\cite{Agakishiev:2011dc} and PHENIX~\cite{Adare:2010pt}; right panel: numerical calculation of the $\phi$ and $\rho^0$ productions in $0-5\%$ $\rm Pb+Pb$ collisions at LHC $2.76$~TeV comparing with preliminary ALICE data from~\cite{Richer:2015vqa}.}
\label{fig:illustphilhc}
\end{center}
\end{figure}

Applying the same calculation framework of $\pi^0$ and $\eta$, we derive the production of the $\rho^{0}$ and $\phi$ both in RHIC and LHC~\cite{Dai:2016}. As we compared the leading order results with the experimental data, we find in the $0-10\%$ most central $\rm Au+Au$ collisions at RHIC $200$~GeV, the nuclear modification factor of $\rho^{0}$ at $\hat q_{0} \tau_{0}=0.72$~GeV$^2$ can explain the data fairly well at large $p_T$ region in the left panel of Fig.~\ref{fig:illustphilhc} .  We also notice that the nuclear modification factor of $\phi$ as a function of $p_T$ however is lower than the limited experimental data points, but still around expected value of $0.2$ which is similar as $\pi^0$ and $\eta$.

In $0-5\%$ $\rm Pb+Pb$ collisions at LHC $2.76$~TeV shown in right  panel of Fig.~\ref{fig:illustphilhc} , we predict the $\rho^{0}$ and $\phi$ productions and compared the numerical results of the nuclear modification factor of $R^{\phi}_{\rm AA}$ as a function of final state $p_T$ with the ALICE preliminary measurements~\cite{Richer:2015vqa}.  The nuclear modification of $\phi$  from our theoretical model is moderately higher than the experimental data. 

Unlike the other three mesons that we investigate, $\phi$, is a dominantly strange meson. The suppression pattern of a strange quark dominated meson is largely depend on the  sophisticated structure of the strange quark fragmentation function. The parameterization of the $\phi$ fragmentation function that we employed here certainly needs more experimental data to be constrained.

\hspace{0.7in}
\hspace*{-1.5in}

\section{Conclusions}
\label{Conclusions}

This series of studies is to uniformly investigate all the identified hadron production ratios and their medium suppressions in the framework of pQCD, hence to pave the way to quantitatively understand the nature of the jet quenching effect in the context of all the identified hadron productions.

We thank X. Chen, H. Zhang and E. Wang for helpful discussions. 
This research is supported by the MOST in China under Project Nos. 2014CB845404, 2014DFG02050, and by NSFC of China with Project Nos. 11322546, 11435004, and 11521064.


\nocite{*}
\bibliographystyle{elsarticle-num}
\bibliography{jos}



\end{document}